# Direct Measurement of Topological Properties of an Exceptional Parabola

Weiyuan Tang[1], Kun Ding[2]*, Guancong Ma[1]*

[1] *Department of Physics, Hong Kong Baptist University, Kowloon Tong, Hong Kong, China*

[2] *Department of Physics, State Key Laboratory of Surface Physics, and Key Laboratory of Micro and Nano Photonic Structures (Ministry of Education), Fudan University, Shanghai 200438, China*

*Email: kunding@fudan.edu.cn, phgcma@hkbu.edu.hk.

**Abstract.** Non-Hermitian systems can produce branch singularities known as exceptional points (EPs). Different from singularities in Hermitian systems, the topological properties of an EP can involve either the winding of eigenvalues that produces a discriminant number (DN) or the eigenvector holonomy that generates a Berry phase. The multiplicity of topological invariants also makes non-Hermitian topology richer than its Hermitian counterpart. Here, we study a parabola-shaped trajectory formed by EPs with both theory and acoustic experiments. By obtaining both the DNs and Berry phases through the measurement of eigenvalues and eigenfunctions, we show that the EP trajectory endows the parameter space with a non-trivial fundamental group consisting of two non-homotopic classes of loops. Our findings not only shed light on exotic non-Hermitian topology but also provide a route for the experimental characterization of non-Hermitian topological invariants.

*Introduction.* The study of non-Hermitian systems [1–3] followed by their realization [4–6] have led to a myriad of intriguing phenomena, such as lasing [7–9], coherent absorption [10–12], scattering cancellation [13,14], sensors [15,16], power transfer devices [17–19], etc. These discoveries are underpinned by the properties of EPs [20,21]. An EP is a branch-point singularity on the Riemannian manifold, which possesses intriguing topological properties that have no counterpart in Hermitian systems. Most notably, unlike the topology in the Hermitian band theory that associates solely with eigenvectors, holonomy of both eigenvectors and eigenvalues around an EP can generate non-Hermitian topological invariants [28–30]. For example, eigenvector holonomy [22] can be extended for non-Hermitian systems [23,24], giving rise to a quantized Berry phase of $\pi$ after two cycles [24–27]. However, eigenvalue holonomy around an EP generates another topological invariant, the DN [28–32].

The fundamental importance of both eigenvalues and eigenvectors near an EP mentioned above implies that their experimental characterizations are crucial for a holistic understanding of non-Hermitian topology. However, most attempts so far mainly focused on the study of state exchanges associated with windings around an isolated EP [26,27]. Recent advances show that even more exotic topology can emerge when considering higher-order EPs [33–35]. In this work, we study a trajectory of EPs in a two-state system with both theory and acoustic experiments. Through a thorough investigation of the topological properties via both DNs and Berry phases, we show that the system's Riemannian manifold is characterized by a non-trivial fundamental group. The system is realized in acoustic experiments. We develop experimental methods relying solely on stationary-state measurement to produce both DNs and Berry phases. The methods avoid any non-adiabatic transition that is commonly encountered in dynamic pumping processes [17,36–38], thus can benefit future studies on non-Hermitian topology.

*A parabola of exceptional points.* —We begin by considering a two-state non-Hermitian Hamiltonian

$$H = (\omega_0 - i\gamma_0)\sigma_0 + H_{EP}, \quad (1)$$

where the first term describes the constant frequency offset and dissipation for each site, with $\omega_0 - i\gamma_0$ being the onsite complex eigenvalue and $\sigma_0$ being an identity matrix. The second term $H_{EP}$ is

$$H_{EP} = -\kappa_0(1 - \zeta^2)\sigma_1 + i\kappa_0(1 + \Xi)(\sigma_3 - \sigma_0). \quad (2)$$

$\sigma_j$ is the $j$-component of the Pauli matrices of spin-1/2. The term $-\kappa_0(1 - \zeta^2)$ ($\kappa_0 > 0$) represents the hopping between the two sites, and it has a quadratic dependence on $\zeta$. The non-Hermicity is induced by a complex parameter $\Xi = \xi_r + i\xi_i$, and $\zeta, \xi_r, \xi_i \in \mathbb{R}$ constitute the three-dimensional (3D) parameter space, as shown in Fig. 1(a). The eigenvalues of Eq. (1) are

$$\omega_{1,2} = \omega_0 - i\gamma_0 - i\kappa_0(1 + \Xi) \pm \kappa_0\sqrt{(1 - \zeta^2)^2 - (1 + \Xi)^2}. \quad (3)$$

We focus on small values of $\zeta, \xi_r, \xi_i$, and thus the eigenvalues become identical when

$$\zeta^2 + \xi_r + i\xi_i = 0, \quad (4)$$

indicating the emergence of EPs. Equation (4) implies $\zeta^2 + \xi_r = 0$ and $\xi_i = 0$, which means the trajectory of EPs forms a parabola on $\xi_r\zeta$-plane, as shown in Fig. 1(a). We denote this EP trajectory as an "exceptional parabola" (EPB).

The EPB invokes intriguing geometry on the system's manifold. We begin by considering its vertex point at $(\xi_r, \xi_i, \zeta) = (0, 0, 0)$, as marked by the red star in Fig. 1(a). It has been shown that this EP is anisotropic in the $\xi_r\zeta$-plane, displaying distinct singular behaviors when approached from $\zeta$ and $\xi_r$-directions [39,40]. Here, the EPB resides in a 3D space, making it possible to define two inequivalent paths that encircle the vertex EP (VEP), as shown in Fig. 1(a-c). Loop-A is on $\xi_i\xi_r$-plane, which is punctured by the EPB once, whereas Loop-B on $\xi_i\zeta$-plane is tangential to the EPB at its vertex. Figure 1(b, c) shows the real parts of eigenvalues on the $\xi_i\xi_r$ and $\xi_i\zeta$-plane respectively, in which eigenvalue evolutions along Loop-A and B are different and thus indicate their topology should have crucial distinctions, which will be discussed later.

*Experimental setup.* —The Hamiltonian can be realized with two coupled acoustic cavities [34,39], as shown in Fig. 1(d). Two identical cuboid cavities with a square cross-section (side length 44 mm) and a height $h = 110$ mm are filled with air. We use the second-order cavity mode as the onsite orbital, which has two nodes in the $z$-direction and is constant in the $xy$-plane. The cavities are horizontally connected by small holes introducing evanescent coupling. The resultant eigenmodes are shown in Fig. 1(e), in which the states 1 and 2 are respectively symmetric and anti-symmetric. To encircle the EP, we need to tune all three parameters in Eq. (2). The term $i\kappa_0(1+\Xi)(\sigma_3-\sigma_0)$ modifies the onsite eigenfrequency and the loss of cavity B. $\xi_r$ tunes dissipation, which is realized by adding acoustic sponges to cavity B. $\xi_i$ introduces detuning and is implemented by adjusting the cavity's volume by the inclusion of putty. To realize $\zeta$ that gives the hopping a quadratic dependence, we leverage on the cavity mode's symmetry by noting that the hopping strength is proportional to the local acoustic intensity $I \propto \cos^2\left(\frac{2\pi z}{h}\right) \cong 1 - \left(\frac{2\pi z}{h}\right)^2$, where $z=0$ is the mirror plane of the cavities. The hopping strength is tunable by adjusting the location of the holes and has a near-quadratic dependence on $z$, thus realizes $\zeta$ [40]. To achieve this, we drill an array of 23×3 holes on a coupling plate. When only the six holes (two rows) near $z=0$ are open (Fig. 1(d)), the hopping is maximum with $\kappa_0 = 48.5$ rad/s, because the local acoustic intensity is maximum. This corresponds to $\zeta = 0$. Other values of $\zeta$ are then realized by opening holes at different $z$.

A small port with a radius of 2 mm is opened at the top of both cavities for the excitation by a loudspeaker placed on cavity A. They also introduce radiative losses, which contribute to $\gamma_0$. The loudspeaker is driven by a waveform generator (Keysight 33500B). We then insert ¼-inch microphones to the side of the cavities at 14 equally spaced ports (7 for each cavity) for the measurement of spectral responses and acoustic fields inside. The signal is recorded by an oscilloscope (Keysight DSO2024A).

*DN and its Measurement.* The eigenvalue Riemannian surfaces intersect at the branch cut(s) (Fig. 1(b, c)), leading to a nonzero eigenvalue vorticity

$$\nu_{jj'} = -\frac{1}{2\pi}\oint_C \nabla_{\vec{\lambda}}\{\arg[\Delta\omega_{jj'}(\vec{\lambda})]\} \cdot d\vec{\lambda}. \tag{5}$$

Here, $\Delta\omega_{jj'}(\vec{\lambda}) = \omega_j(\vec{\lambda}) - \omega_{j'}(\vec{\lambda})$ wherein $\omega_j(\vec{\lambda})$ is the complex eigenvalue of the $j$th state and $j \neq j'$. This integral operates on the phase gradient of $\Delta\omega_{jj'}(\vec{\lambda})$ over a closed path $C$ with $\vec{\lambda}$ being the relevant parameters. The Stoke's theorem tells the integral must vanish unless the region enclosed by $C$ contains indifferentiable point(s), i.e., singularities. Equation (5) can be computed stepwise

$$\nu_{jj'} = -\frac{1}{2\pi}\sum_{l=1}^{\mathcal{L}} \text{Im}\left\{\ln\left[\frac{\Delta\omega_{jj'}(\lambda_{l+1})}{\Delta\omega_{jj'}(\lambda_l)}\right]\right\}, \tag{6}$$

where $\lambda_l$ is the parametric coordinate of the $l$-th step on $C$. The DN is formed by summing over the vorticities [30,41]. For our two-state system, the DN is $\mathcal{V} = \nu_{12} + \nu_{21}$.

We can obtain $\mathcal{V}$ experimentally by measuring the eigenvalues $\omega_{1,2}$ [40]. We tune the acoustic system along an encircling path (see parameters in [40]). Two microphones are used to measure the pressure response spectra at each parametric point. The eigenvalues are then retrieved using a Green's function method [34,40]. The results of Loop-A are shown in Fig.

2(a, b). Excellent agreement with theory is achieved. We see that the $\Delta\omega_{12}$ and $\Delta\omega_{21}$ form a closed loop on the complex frequency plane, as shown by the dashed curves in Fig. 2(a). This case is a point gap, which has no Hermitian counterpart [30,42]. Both arguments of $\Delta\omega_{12}$ and $\Delta\omega_{21}$ accumulate a $\pi$ phase after one complete cycle (Fig. 2(b)), indicating $\nu_{12} = \nu_{21} = -0.5$. The DN is thus $\mathcal{V}_{\xi_i\xi_r} = -1$. In contrast, when following Loop-B in Fig. 1(a), the eigenvalue differences return to the respective starting point without forming a closed loop (Fig. 2(c, d)). The DN is hence $\mathcal{V}_{\xi_i\zeta} = 0$, which is different from $\mathcal{V}_{\xi_i\xi_r}$ and topologically trivial. This is because the two states form a line gap on the complex frequency plane (Fig. 2(c)), which can be smoothly transformed into Hermitian bands [30].

Although both loops enclose the VEP, two different DNs are obtained. It follows that through the winding of eigenvalues, we can readily see that the two loops must belong to different equivalence classes. Consequently, the fundamental group of the system's manifold is non-trivial.

*Measurement of the Berry phase.* The cyclic variations also drive the eigenvectors around the VEP, which give rise to a Berry phase

$$\Theta = i \oint_C d\vec{\lambda} \cdot \langle \psi_j^L(\vec{\lambda}) | \nabla_{\vec{\lambda}} | \psi_j^R(\vec{\lambda}) \rangle, \tag{7}$$

with $|\psi_j^R(\vec{\lambda})\rangle$ and $\langle \psi_j^L(\vec{\lambda})|$ being the right and left eigenvectors. Likewise, Eq. (7) can be discretized

$$\Theta = -\sum_{l=1}^{N\mathcal{L}} \text{Im}\left[\ln\langle \psi_j^L(\lambda_l) | \psi_j^R(\lambda_{l+1}) \rangle\right], \tag{8}$$

where $N$ is the number of cycles for all states to recover. Equation (8) not only suits numerical computations [43] but also offers a viable path to experimentally obtain $\Theta$. Although the Berry phase emerges from an adiabatic process, the result has no explicit dependence of time. This removes the need for a dynamic adiabatic process, and all information can be obtained by measuring the stationary-state eigenfunctions [40]. However, because stationary-state measurement involves the independent excitation of the system at different $\lambda_l$, the results are inevitably contaminated by arbitrary phase factors. To eliminate such arbitrary phases, a specific gauge is applied to guarantee the parallel transport at each step [44]. This method also has the advantage of visualizing the accumulation of Berry phase.

In practice, we first experimentally obtain the eigenfunctions by measuring the field distribution in both cavities at a total of 14 positions (Fig. 1(d)), from which the eigenfunctions are obtained [40]. The arbitrary phases can be identified by $\beta_j(\lambda_{l+1}) = \text{Im}\left[\ln\langle \psi_j^L(\lambda_l) | \psi_j^R(\lambda_{l+1}) \rangle\right]$. $\beta_j(\lambda_{l+1})$ and then removed by $|\bar{\psi}_j^R(\lambda_{l+1})\rangle = e^{-i\beta_j(\lambda_{l+1})} |\psi_j^R(\lambda_{l+1})\rangle$. For the starting point, we set $|\bar{\psi}_j^R(\lambda_1)\rangle = |\psi_j^R(\lambda_1)\rangle$, meaning that the

initial phase serves as the chosen constant gauge. Repeating this procedure for all the following steps eliminates any arbitrary phases, thus ensures all neighboring eigenfunctions are locally parallel along the path [40].

The evolution of eigenfunctions along Loop-A is shown in Fig. 3(a). We follow the state with two cavities being out-of-phase at the starting point (Point I-1). Mode exchange occurs near Point I-2, at which the field in cavity B nearly vanishes, implying the mixing of the two states. Starting from Point I-3, the two cavities are in-phase until the completion of one cycle. Then, the second cycle begins with a state (II-1) that is almost identical with the state at I-8 (Fig. 3(a)), followed by a mode switching (II-2) across which the two cavities become out-of-phase and remain until the end of the cycle. Upon the completion of two cycles, the state is restored. However, the two states at I-1 and II-8 clearly have a phase difference of $\pi$, which is the Berry phase. This is further confirmed in Fig. 3(b), showing the accumulated Berry phase is $\Theta_{\xi_i\xi_r} = -\pi$ for two cycles. In contrast, for Loop-B, the eigenfunction exchanges twice (near I-2 and I-6) and are fully restored in only one complete cycle (Fig. 3(c)). The Berry phase is $\Theta_{\xi_i\zeta} = 0$ (Fig. 3(d)). Although the values of $\beta_j(\lambda_l)$ are path-dependent, the holonomic Berry phases $\Theta$ are gauge-invariant when the states are restored. To further verify, we employ the Wilson-loop method [45], which gives $\Theta_{\xi_i\xi_r} = -0.9971\pi$ and $\Theta_{\xi_i\zeta} = 1.7 \times 10^{-4}$, conforming welling with the parallel-transport results.

*Discussion and conclusion.* An order-2 EP carries a "topological half charge" [46] owing to a Berry phase of $\pm\pi$ after two complete cycles [24–27,47]. Our results unveil that this is not always the complete picture. In fact, Loops-A and B belong to two different equivalence classes in a non-trivial fundamental group. We confirm this by examining homotopic loops. For example, Loop-A can be transformed smoothly to Loop-A' (Fig. 1(a)), and both loops produce the same DN and Berry phase, which we have confirmed with both theory and experiments [40]. Likewise, Loops-B and B' are also homotopic, despite that Loop-B encircles only the VEP but B' encircles two EPs of the same order.

Further insights about the loops' homotopic equivalence can be obtained by the discriminant fields $\vec{D}(\vec{\lambda}) = \nabla_{\vec{\lambda}}[\text{Im}(\ln \Delta)]$, where $\Delta$ is the discriminant of $\det(\omega\sigma_0 - H_{EP}) = 0$. Figure 4 shows the computed results. A vortex centering at the VEP is seen on $\xi_i\xi_r$-plane but not on the $\xi_i\zeta$-plane. Loop-B hence encloses zero vorticity flux, although the singularity (the VEP) is still at the center. At $\xi_r = -0.24$, $\xi_i\zeta$-plane cuts both arms of the EPB and each intersecting EP carries a vortex with opposite chirality, leading to a zero net vorticity (Fig. 4(c)), which is equivalent to Fig. 4(b).

Our work forms a coherent and holistic picture of the topological properties of the non-Hermitian Riemannian manifold. The DN is a topological invariant that associates with the eigenvalue holonomy and has no non-trivial Hermitian counterparts since it always vanishes for real eigenvalues. The DN highlights its role in the following senses [40]. Firstly, it is an indicator distinguishing point gaps from line gaps. To further demonstrate, we present a relevant example in which the VEP is a chain point and Loop-B yields a non-trivial DN, in which case the line gap shown in Fig. 2(c) becomes a point gap [40]. Recent works showed that the DN was also linked to the non-Hermitian skin effect [48,49], thus generalizing bulk-edge correspondence to non-Hermitian scenarios. On the other hand, the Berry phase results from eigenvector holonomy on the Riemannian manifold [23,24], making it an entirely different invariant from DN. Due to the possibility of mode exchange, the non-Hermitian Berry phase can be quantized after one or more complete cycles, giving rise to the possibility of fractional winding numbers [35,46], while the Hermitian counterparts are always integers. In sum, our investigation shows that Berry phase and DN produce coherent outcomes revealing signatures of non-Hermitian topology.

Lastly, our experimental approaches involve only stationary-state measurements and are immune to non-adiabatic transitions that are almost inevitable in non-Hermitian dynamic pumping [17,18,36,37]. Our methods thus offer robust routes to directly characterize topological invariants without relying on any descendant phenomena such as momentum transfer [50], polarization rotation [46], or bulk-surface correspondence [51].

*Acknowledgments.* We thank Ruoyang Zhang and Xiaohan Cui for fruitful discussions. This work was supported by Hong Kong Research Grants Council (12302420, 12300419, 22302718, C6013-18G), National Natural Science Foundation of China (11922416,11802256), and Hong Kong Baptist University (RC-SGT2/18-19/SCI/006).

(2012).

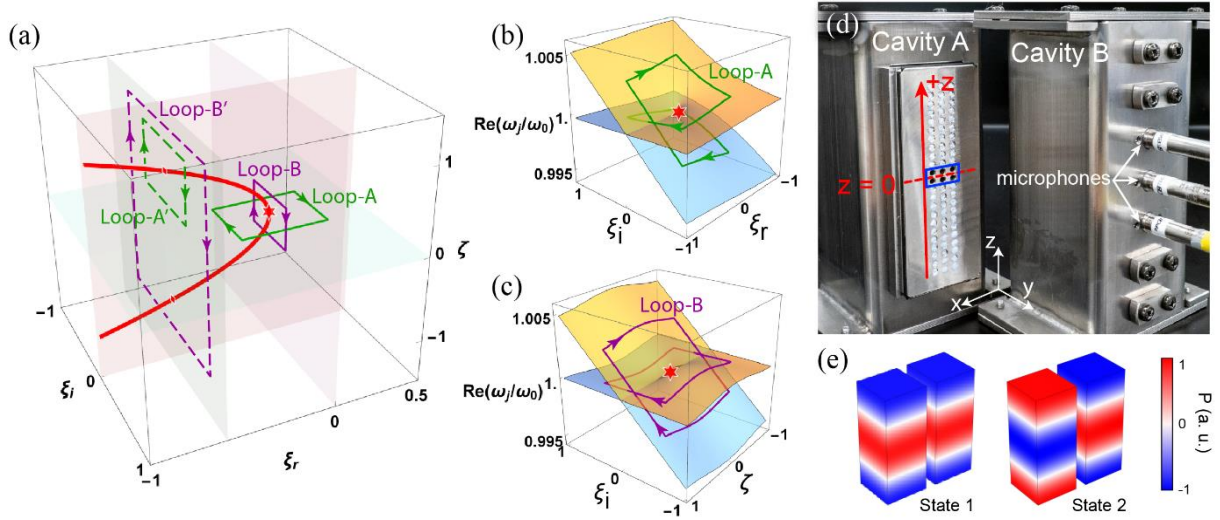

FIG. 1. (a) An EPB (solid red curve) lies on the $\xi_r\zeta$-plane ($\xi_i = 0$). The VEP is marked by the red star. The green and purple boxes denote encircling paths. (b, c) show the real parts of eigenvalues near the VEP on the $\xi_i\xi_r$ and $\xi_i\zeta$-plane. The paths herein indicate the eigenvalue evolutions along the Loop-A and B in (a). The eigenvalues are normalized by $\omega_0 = 19{,}613$ rad/s. (d) The experimental setup. (e) Acoustic modes with non-Hermicity absent (($\xi_r, \xi_i, \zeta) = (-1,0,0)$).

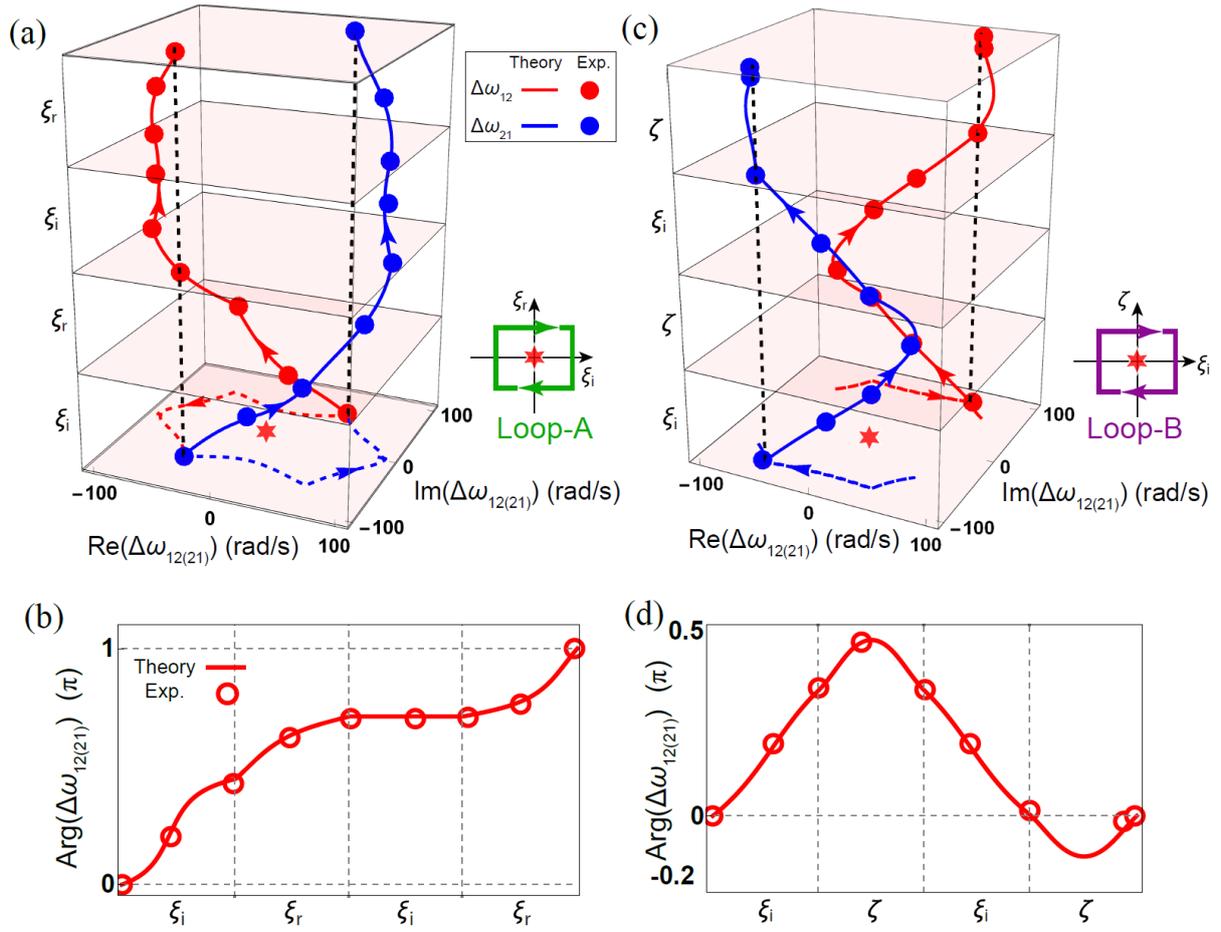

FIG. 2. The evolution of eigenvalue differences $\Delta\omega_{12}$ and $\Delta\omega_{21}$ and their argument changes along Loop-A and B are shown in (a, b) and (c, d), respectively. In (b, d), the argument of $\Delta\omega_{21}$ are identical to $\Delta\omega_{12}$. Estimated errors are smaller than the marker sizes.

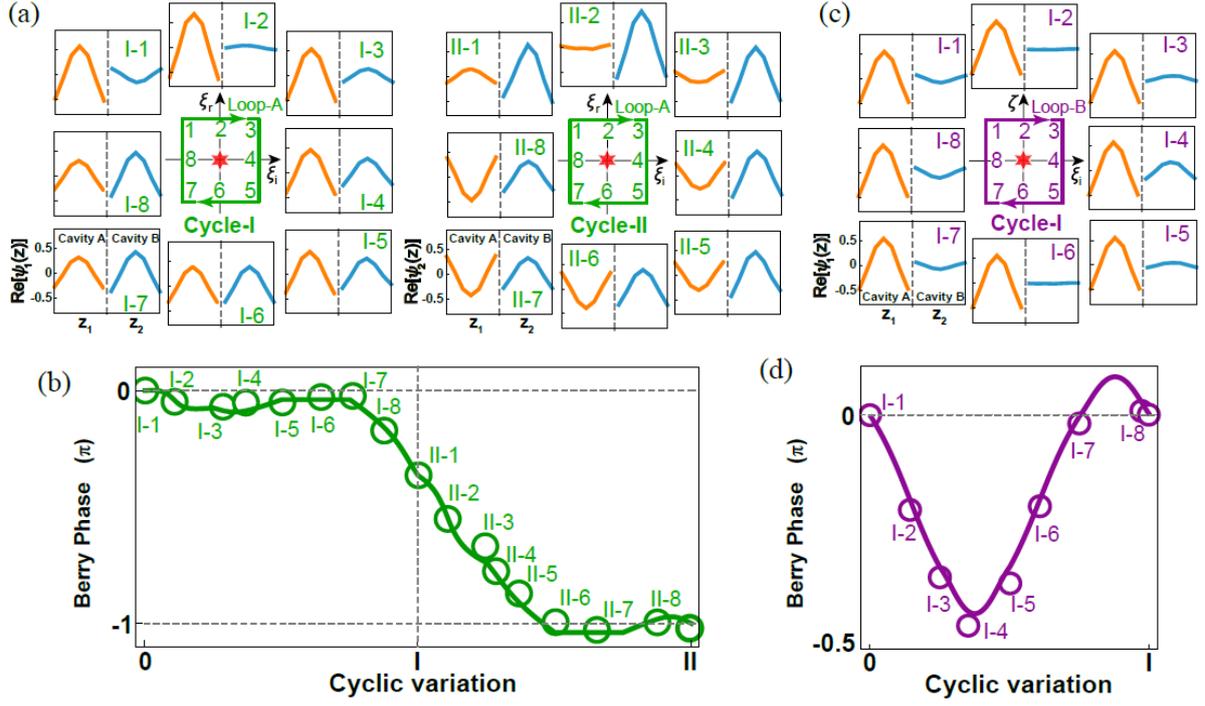

FIG. 3. (a) Measured eigenfunctions (real parts) along Loop-A. The orange and cyan lines respectively denote cavity A and B. (b) The corresponding Berry phase is $-\pi$ after two cycles. (c) Along Loop-B, the measured eigenfunctions exchange twice near Points I-2 and I-6 and restore themselves in one cycle. (d) The Berry phase is 0. The markers and lines in (b, d) are experimental and theoretical results, respectively. Estimated errors are smaller than the marker size in (b, d).

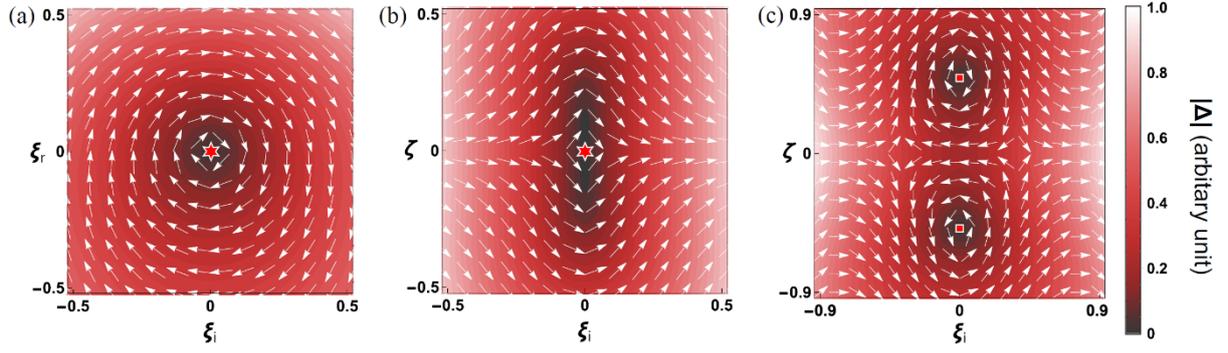

FIG. 4. The calculated discriminant fields $\vec{D}$ (arrows) and its norm (color map) near the VEP (red stars) on the (a) $\xi_r\xi_i$-plane and (b) $\xi_i\zeta$-plane. (c) When $\xi_r = -0.24$, $\xi_i\zeta$-plane intersects with the EPB twice at the red squares, around which two vortices with opposite chirality are seen. Arrow length is normalized by $|\vec{D}|$.

# Supplemental Materials

## "Direct Measurement of Topological Properties of an Exceptional Parabola"

Weiyuan Tang[1], Kun Ding[2], Guancong Ma[1]

[1] *Department of Physics, Hong Kong Baptist University, Kowloon Tong, Hong Kong, China*

[2] *Department of Physics, State Key Laboratory of Surface Physics, and Key Laboratory of Micro and Nano Photonic Structures (Ministry of Education), Fudan University, Shanghai 200438, China*

**Contents**

## I. Trajectory of EPs

The eigenvalues of the Hamiltonian $H_{EP}$ defined in Eq. (2) of the main text are

$$\omega_{1,2} = \omega_0 - i\gamma_0 - i\kappa_0(1+\Xi) \pm \kappa_0\sqrt{(1-\zeta^2)^2 - (1+\Xi)^2}. \tag{1}$$

In the main text, we focus on the region that all three parameters are smaller than 1, so that the higher-order terms under the square root, namely the discriminant, are dropped. Here, we examine the trajectory of the EPs by retaining all the terms. The EPs are found when

$$\zeta^4 - 2\zeta^2 - \xi_r^2 + \xi_i^2 - 2i\xi_i\xi_r - 2\xi_r - 2i\xi_i = 0, \tag{2}$$

which further indicates

$$\zeta^4 - 2\zeta^2 - \xi_r^2 + \xi_i^2 - 2\xi_r = 0, \tag{3}$$

$$\xi_i\xi_r + \xi_i = 0. \tag{4}$$

These two equations give the EP trajectories, which are two exceptional parabolas (EPBs) on the $\xi_r\zeta$-plane with $\xi_i = 0$ and are symmetric about $\xi_r = -1$ (see Fig. S1). Note that these two EPBs intersect at two diabolic points at $(\xi_r, \xi_i, \zeta) = (-1, 0, -1)$ and $(-1, 0, 1)$, respectively, which corresponds to the Hermitian case.

Since the EPs on the trajectory have more than one parameter dependence, they possess distinct singular behaviors when approached from different parametric directions [1]. For example, for the vertex EP (VEP) at $(\xi_r, \xi_i, \zeta) = (0,0,0)$. When the EP is approached along the $\xi_r$ direction with $\zeta = 0$ and $\xi_i = 0$, the eigenvalues produce a square-root singularity (red dots in Fig. S2 (a, b)). When approached from $\xi_i$ with $\zeta = \xi_r = 0$, the splitting is also square root, but the real parts split for both $\xi_i > 0$ and $\xi_i < 0$ (green dots in Fig. S2(a)). However, along $\zeta$ axis with $\xi_r = \xi_i = 0$ the real eigenvalue does not respond to $\zeta$ at all (blue dots in Fig. S2(a)), whereas the imaginary parts follow a linear function of $\zeta$ in the vicinity of $\zeta = 0$ (blue dots in Fig. S2(b)). These different splitting behaviors in eigenvalues can be leveraged for applications. For example, a sensor operating near such an EP can respond to different types of perturbations with different sensitivities, granting it the potential to monitor different variables simultaneously.

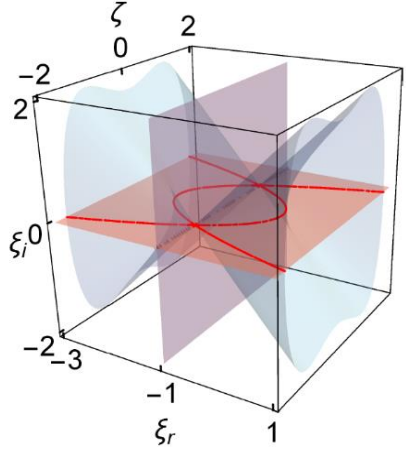

FIG. S1. Equation (1) yields two EPBs (solid red lines) on the $\xi_r\zeta$-plane (with $\xi_i = 0$), which are symmetric about $\xi_i\zeta$-plane with $\xi_r = -1$. The blue surfaces correspond to the solutions of Eq. (3), and the red planes are the solutions to Eq. (4).

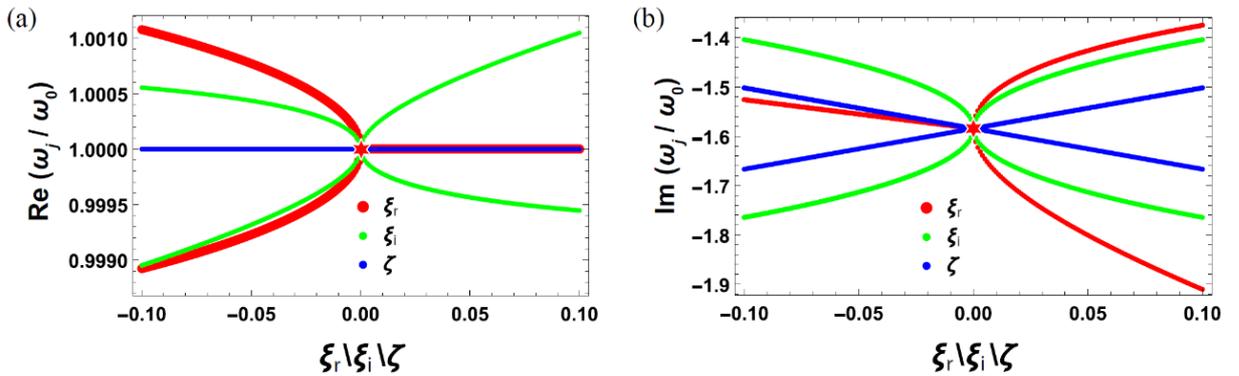

FIG. S2. Eigenvalue splitting near the VEP (red stars) follows distinct behaviors along different parametric directions. (a) and (b) show the splitting of the real and imaginary parts, respectively.

## II. Imaginary parts of the eigenvalue Riemannian surfaces near the vertex EP

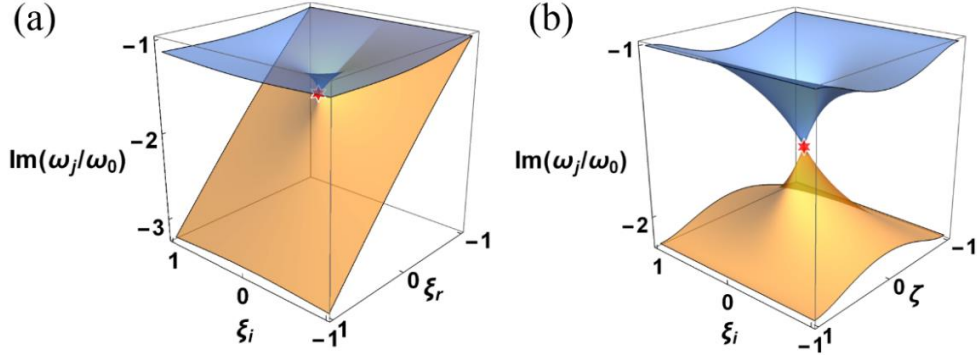

FIG. S3. Imaginary parts of eigenvalues (a) on $\xi_r\xi_i$ plane, and (b) on $\xi_i\zeta$ plane. The red stars indicate the exceptional point at the parabola's vertex. The parameters here are the same as those in Figs. 1(b) and 1(c), which show the real parts of eigenvalues.

## III. Realization the linear and quadratic dependence of parameters.

The realization of the EPB requires the loss to have a linear dependence on $\xi_r$, and the hopping strength to have a quadratic dependence on $\zeta$. The loss is achieved by inserting small pieces of acoustic sponge into the chosen cavity. We put all the sponge pieces to the bottom of the cavity so that they all experience the same acoustic field. Thus the amount of sponge directly tunes the total loss. In the benchmark results shown in Fig. S4(a), we can see that the loss is linearly proportional to the volume of acoustic sponges. Hence, $\xi_r$ is implemented as the sponge volume in our experiments. On the other hand, as mentioned in the main text, the quadratic dependence of hopping on $\zeta$ is realized by exploiting the cosine acoustic field distribution along the z-direction (Fig. 1(e), main text). The acoustic field has an anti-node in the middle plane. There, in the vicinity of the middle plane, whose position is chosen as $z = 0$, the acoustic intensity is $I(z) \propto \cos^2\left(\frac{2\pi z}{h}\right) \cong 1 - \left(\frac{2\pi z}{h}\right)^2 \sim 1 - \zeta^2$. In other words, the role of $\zeta$ is played by $2\pi z/h$. In Fig. S4(b), we plot the hopping coefficient as a function of $z$. A quadratic dependence is clearly seen near $z = 0$.

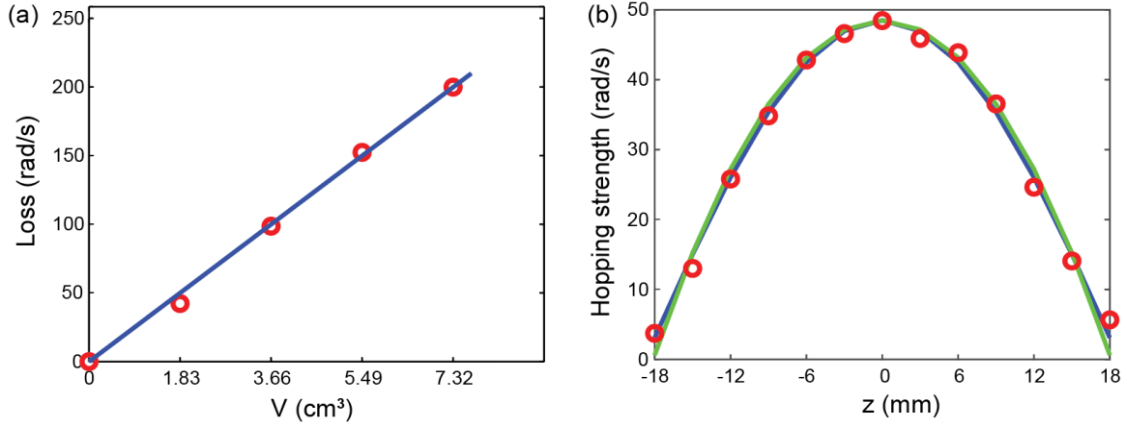

FIG. S4. (a) The linear dependence of loss on the volume of the acoustic sponge. (b) The hopping coefficient as a function of $z$, in which a quadratic dependence is seen near $z = 0$. The blue and green curves are cosine and quadratic fit, respectively. Red circles in (a) and (b) are experimentally measured values.

## IV. Method for obtaining the DN

The DN is a non-Hermitian topological invariant defined solely on eigenvalues. To calculate the DN, the first step is to obtain the vorticity of the complex eigenvalues

$$v_{jj'} = -\frac{1}{2\pi}\oint_{\partial\Omega} \nabla_{\vec{\lambda}} \arg[\Delta\omega_{jj'}(\vec{\lambda})] \cdot d\vec{\lambda}, \tag{5}$$

which is evaluated on a closed loop $\partial\Omega(\vec{\lambda})$ defined on a manifold $\Omega$, for example, the closed Loop-A and B in Fig. 1(a) [2,3]. Here, $\Delta\omega_{jj'}(\vec{\lambda}) = \omega_j(\vec{\lambda}) - \omega_{j'}(\vec{\lambda})$ is the difference of the complex eigenvalues of two different states labeled by $j$ and $j'$ $(j \neq j')$. Equation (5) is a closed-loop integral with the gradient field of the argument of $\Delta\omega_{jj'}$. It can be transformed to a surface integral using the Stoke's theorem,

$$v_{jj'} = -\frac{1}{2\pi}\iint_\Omega \nabla_{\vec{\lambda}} \times \{\nabla_{\vec{\lambda}} \arg[\Delta\omega_{jj'}(\vec{\lambda})]\} \cdot d\vec{S}_{\vec{\lambda}}. \tag{6}$$

Note that the integrand now becomes the curl of the gradient field, which must vanish as long as the manifold $\Omega$ is differentiable at every point over the integrated area. Equivalently, this means for $v_{jj'}$ to have nonzero values, $\Omega$ must contain one or more singularities. In the context of Riemannian manifolds of non-Hermitian systems, such singularities are obviously EPs.

Equation (5) can be calculated by discretizing the loop $\partial\Omega$ into $\mathcal{L}$ equally spaced small segments with each step being $\Delta\lambda$

$$v_{jj'} = -\frac{1}{2\pi}\left\{\frac{\arg[\Delta\omega_{jj'}(\lambda_2)] - \arg[\Delta\omega_{jj'}(\lambda_1)]}{\Delta\lambda} + \cdots \frac{\arg[\Delta\omega_{jj'}(\lambda_{l+1})] - \arg[\Delta\omega_{jj'}(\lambda_l)]}{\Delta\lambda}\right.$$
$$\left. + \cdots \frac{\arg[\Delta\omega_{jj'}(\lambda_{\mathcal{L}})] - \arg[\Delta\omega_{jj'}(\lambda_{\mathcal{L}-1})]}{\Delta\lambda}\right\}\Delta\lambda,$$

and then leads to

$$v_{jj'} = -\frac{1}{2\pi}\sum_{l=1}^{\mathcal{L}}\{\arg[\Delta\omega_{jj'}(\lambda_{l+1})] - \arg[\Delta\omega_{jj'}(\lambda_l)]\}$$
$$= -\frac{1}{2\pi}\sum_{l=1}^{\mathcal{L}} \text{Im}\left\{\ln\left[\frac{\Delta\omega_{jj'}(\lambda_{l+1})}{\Delta\omega_{jj'}(\lambda_l)}\right]\right\}, \quad (7)$$

where $\lambda_{l+1} = \lambda_l + \Delta\lambda$ is the discretized parametric coordinate of the $l$-th step on a closed-loop, and $\mathcal{L}$ is the total number of steps that closes the loop. Note that the arguments produced in computer programs are typically the principal values, *i.e.*, $\text{mod}(2\pi)$, so that in practice, one cannot simplify Eq. (7) by taking the difference of $\arg(\Delta\omega_{jj'})$ between the final and the first steps.

The DN is then the summation of vorticity $v_{jj'}$ for all combinations of $j, j'$,

$$\mathcal{V} = \sum_{j\neq j'} v_{jj'}. \quad (8)$$

For our two-state system, the DN is simply $\mathcal{V} = v_{12} + v_{21}$.

It is noteworthy that an equivalent expression of the DN is also a closed-loop integral of the discriminant fields $\vec{D}(\vec{\lambda}) = \nabla_{\vec{\lambda}}\{\text{Im}[\ln(\Delta)]\}$, where $\Delta$ is the discriminant of the characteristic polynomial of the Hamiltonian. Thus the discriminant fields $\vec{D}(\vec{\lambda})$ can also reveal the topology in our system, as seen in Fig. 4 in the main text.

## V. Retrieval of eigenvalues

The DNs can be obtained from experimental data by retrieving the eigenvalues. We employ the Green's function method [1,4] to obtain the eigenvalues from the measured response spectra. The Green's function is defined as

$$\overleftrightarrow{G}(\omega, \lambda_l) = \sum_{j=1}^{2} \frac{|\psi_j^R(\lambda_l)\rangle\langle\psi_j^L(\lambda_l)|}{\omega - \omega_j}, \quad (9)$$

where $\omega_j$ denote the eigenvalues of the Hamiltonian with $j = 1, 2$ labeling the eigenstates. $|\psi_j^R(\lambda_l)\rangle$ and $\langle\psi_j^L(\lambda_l)|$ are the normalized biorthogonal right and left eigenvectors, respectively. Note that $\langle\psi_j^L(\lambda_l)| = |\psi_j^R(\lambda_l)\rangle^T$ because of reciprocity. $\lambda_l$ represents the

parametric coordinate of the $l$-th step along a path, which in our work here, is a loop encircling an EP. Then the pressure responses measured at a specific parameter point $\lambda_l$ inside the two cavities are given by

$$P(\omega, \lambda_l) = \langle m | \vec{G}(\omega, \lambda_l) | s \rangle, \qquad (10)$$

wherein $|s\rangle$ and $|m\rangle$ respectively represent the source and microphone information. For the representation implied by Eq. (5) here, $|s\rangle$ and $|m\rangle$ are 2×1 column vectors. In our experiment, we excite at the top of cavity A and measure at the central position of both cavities, so $|s\rangle = (1 \quad 0)^T$ and $|m\rangle$ are $(1 \quad 0)^T$ and $(0 \quad 1)^T$ for measurement at cavity A and B, respectively. By applying a least-square fitting to the amplitude of the pressure responses, we can obtain all system parameters: $\omega_0$, $\gamma_0$, $\kappa_0$, $\xi_r$, $\xi_i$, $\zeta$.

## VI. Methods for obtaining the Berry phase

The multiband Berry phase is defined as

$$\Theta = i \oint_{\partial \Omega} d\vec{\lambda} \ \text{Tr} \mathbf{A}, \qquad (11)$$

wherein $A_{jj'} = \langle \psi_j^L(\vec{\lambda}) | \nabla_{\vec{\lambda}} | \psi_{j'}^R(\vec{\lambda}) \rangle$ is the multiband Berry connection matrix. Similar to the computation of the DN, for each discrete step along the encircling path, Eq. (11) can be transformed to

$$\Theta = - \sum_{l=1}^{\mathcal{L}} \text{Im}[\ln(\det \mathbf{M})]. \qquad (12)$$

where $M_{jj'} = \langle \psi_j^L(\lambda_l) | \psi_{j'}^R(\lambda_{l+1}) \rangle$. Equation (12) is equivalent to

$$\Theta = -\text{Im}\{\ln[\prod_{l=1}^{N\mathcal{L}} \langle \psi_j^L(\lambda_l) | \psi_j^R(\lambda_{l+1}) \rangle]\}. \qquad (13)$$

This is the Wilson-loop method [5]. Term by term, Eq. (13) is

$$\Theta = -\text{Im}\left\{\ln\left[\begin{array}{c}\langle \psi_j^L(\lambda_1) | \psi_j^R(\lambda_2) \rangle \langle \psi_j^L(\lambda_2) | \psi_j^R(\lambda_3) \rangle \dots \langle \psi_j^L(\lambda_l) | \psi_j^R(\lambda_{l+1}) \rangle \\ \dots \langle \psi_j^L(\lambda_{N\mathcal{L}-1}) | \psi_j^R(\lambda_1) \rangle\end{array}\right]\right\}. \qquad (14)$$

We remark that Eq. (12) involves multiple states whereas Eqs. (13, 14) trace one state. The latter case describes evolution is on a complex Riemannian manifold, thus it can evolve smoothly into the other state. Therefore, to obtain a quantized multiband Berry phase, a total of $N \geq 1$ cycles may be necessary so that the end state restores the initial state $|\psi_j^R(\lambda_{N\mathcal{L}})\rangle = |\psi_j^R(\lambda_1)\rangle$ at the last step. Therefore, the total number of steps is $N\mathcal{L}$, where $\mathcal{L}$ is the number of steps in one complete cycle. In our system, $N = 2$ for Loop-A, $N = 1$ for Loop-B. By performing a complete holonomy this way, any arbitrary phase factors carried by the eigenvectors at each step automatically cancel because the left and right eigenvectors of each step are bi-orthogonally normalized, i. e., $\langle \psi_j^L(\lambda_l) | \psi_j^R(\lambda_l) \rangle = 1$ for all $l$. The outcome of Eqs.

(12–14) is thus gauge invariant.

Although the Wilson-loop method can conveniently produce a gauge-invariant Berry phase, it cannot trace the Berry phase's evolution in the intermediate steps. In order to achieve the latter, we perform parallel transport of the eigenvectors at each step along the loop [6]. However, experimentally, the system is independently excited at different $\lambda_l$ in the stationary-state measurement, the eigenfunctions obtained from the measured responses inevitably carry random phase factors. In other words, the measured eigenfunctions do not satisfy parallel transport. Therefore, an additional step is needed to remove the random phase factor at each step.

The procedure is as follows. For every intermediate step $\lambda_l$ ($l > 1$), we can identify the arbitrary phase factor by $\beta_j(\lambda_{l+1}) = \text{Im}[\ln\langle\psi_j^L(\lambda_{l+1})|\bar{\psi}_j^R(\lambda_l)\rangle]$. Then $|\psi_j^R(\lambda_{l+1})\rangle$ is simply rotated by $-\beta_j(\lambda_{l+1})$ to get rid of the arbitrary phase, namely,

$$|\bar{\psi}_j^R(\lambda_{l+1})\rangle = e^{-i\beta_j(\lambda_{l+1})}|\psi_j^R(\lambda_{l+1})\rangle. \tag{15}$$

As a result, $\langle\bar{\psi}_j^L(\lambda_{l+1})|\bar{\psi}_j^R(\lambda_l)\rangle$ becomes real and positive, indicating $|\bar{\psi}_j^R(\lambda_{l+1})\rangle$ and $|\bar{\psi}_j^R(\lambda_l)\rangle$ are almost parallel. For the starting point $\lambda_1$, we set $|\bar{\psi}_j^R(\lambda_1)\rangle = |\psi_j^R(\lambda_1)\rangle$, indicating the initial phase factor is carried along in the smooth holonomy. Therefore, the phase change between the initial and final steps is gauge-invariant. This feature shows the method allows us to trace the Berry phase evolution, which leads to the results shown in Fig. 3(b, d) and Figs. S6(b), S7(b).

## VII. Retrieval of eigenfunctions

To obtain the Berry phase, we need to acquire the eigenfunctions from experimental data. Compared to eigenvalue retrieval, this requires additional information and steps. We note that the eigenfunctions can be expressed as a linear combination of the two onsite modes, *i.e.*,

$$|\psi_j^R(\lambda_l)\rangle = \begin{bmatrix} a_{j,A}(\lambda_l)|\varphi_A\rangle \\ a_{j,B}(\lambda_l)|\varphi_B\rangle \end{bmatrix}, \tag{16}$$

$$\langle\psi_j^L(\lambda_l)| = [b_{j,A}(\lambda_l)\langle\varphi_A|, b_{j,B}(\lambda_l)\langle\varphi_B|], \tag{17}$$

where $|\psi_j^R(\lambda_l)\rangle$ and $\langle\psi_j^L(\lambda_l)|$ are the right and left eigenfunctions of the coupled system, and $|\varphi_{A,B}\rangle$ is a column vector representing the normalized onsite eigenmode, i. e., cavity A or B in isolation. Note that $\langle\varphi_{A,B}| = |\varphi_{A,B}\rangle^\dagger$, since for a single cavity, the Hamiltonian is Hermitian. Therefore, we first need to obtain the onsite eigenmodes by measuring the spatial distribution of the acoustic field inside the two isolated cavities at each parameter point. Since we employ a longitudinal mode that resonates along the *z*-axis (Fig. 1(e), main text), we only need to measure both the amplitude and phase at 7 positions along the side boundary of both cavities (Fig. 1(d), main text). The data at a total of 31 frequencies are measured. Then the measured pressure responses are fit using the Green's function method, which gives

$$P_A(\omega) = \frac{\langle m|\varphi_A\rangle\langle\varphi_A|s\rangle}{\omega-(\omega_0-i\gamma_0)}, \tag{18}$$

$$P_B(\omega) = \frac{\langle m|\varphi_B\rangle\langle\varphi_B|s\rangle}{\omega-[\omega_0-i\gamma_0-2i\kappa_0(1+\Xi)]}, \tag{19}$$

wherein $|m\rangle$ and $|s\rangle$ are now 7×1 column vectors, with each element representing the positions of the source (loudspeaker) and the microphone. Note that since the loudspeaker's position is fixed, there is only one nonzero element in $|s\rangle$. The onsite eigenfunctions $|\varphi_{A,B}\rangle$ are then obtained by numerical fitting both the amplitude and phase of the measured acoustic field in each individual cavity.

Next, we experimentally obtain the acoustic fields in the coupled acoustic cavity system at the same 31 frequencies. For the coupled system, the acoustic field is given by

$$P_j(\omega,\lambda_l) = \langle m|\overleftrightarrow{G}(\omega,\lambda_l)|s\rangle = \sum_{j=1}^{2}\frac{\langle m|\psi_j^R(\lambda_l)\rangle\langle\psi_j^L(\lambda_l)|s\rangle}{\omega-\omega_j}. \tag{20}$$

$|m\rangle$ and $|s\rangle$ now become 14×1 column vectors, representing the position of the loudspeaker and microphone. Then, by expressing the eigenfunctions using Eqs. (16, 17), we perform a least-square fitting on the measured data to acquire $a_{j,A}$, $a_{j,B}$, $b_{j,A}$, $b_{j,B}$, thereby obtaining the right and left eigenfunctions $|\psi_j^R(\lambda_l)\rangle$ and $\langle\psi_j^L(\lambda_l)|$. This procedure is repeated for each parametric point $\lambda_l = (\xi_r, \xi_i, \zeta)$ along the encircling loops.

### VIII. DNs and Berry phases from the dashed loops in Fig. 1(a)

In Fig. S5, we show the results of encircling the EPB along the two dashed loops (Fig. 1(a) in the main text). Loop-A' (the dashed green loop) encloses one arm of the EPB, whereas Loop-B' (the dashed purple loop) encloses both arms.

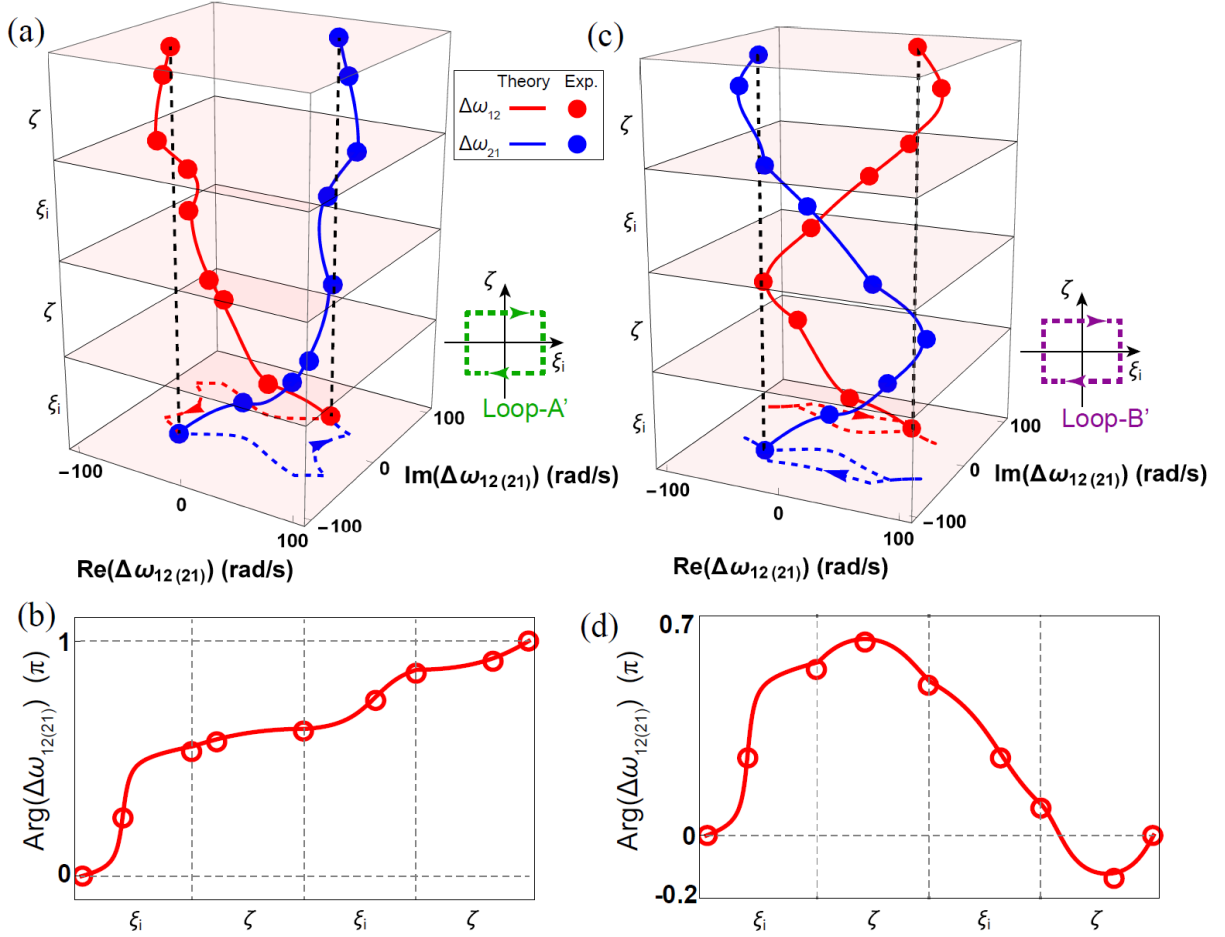

FIG. S5. The evolution of eigenvalue differences $\Delta\omega_{12}$ and $\Delta\omega_{21}$ and their argument changes on the $\xi_i\zeta$-plane with $\xi_r = -0.24$ along the dashed paths in Fig. 1(a) in the main text. (a) Along the dashed green path, the eigenvalue differences $\Delta\omega_{12}$ and $\Delta\omega_{21}$ form a closed loop and each accumulates a phase of $\pi$ (b), which yields a discriminant number of $-1$. (c) Along the dashed purple path, $\Delta\omega_{12}$ and $\Delta\omega_{21}$ return to their starting points in one complete circle and the accumulated phase is 0 (d). The discriminant number is hence 0. The markers are experimental results, and the curves are from theory. In (b, d), $\Delta\omega_{12}$ and $\Delta\omega_{21}$ are identical, therefore the blue curves and markers are covered by the red ones.

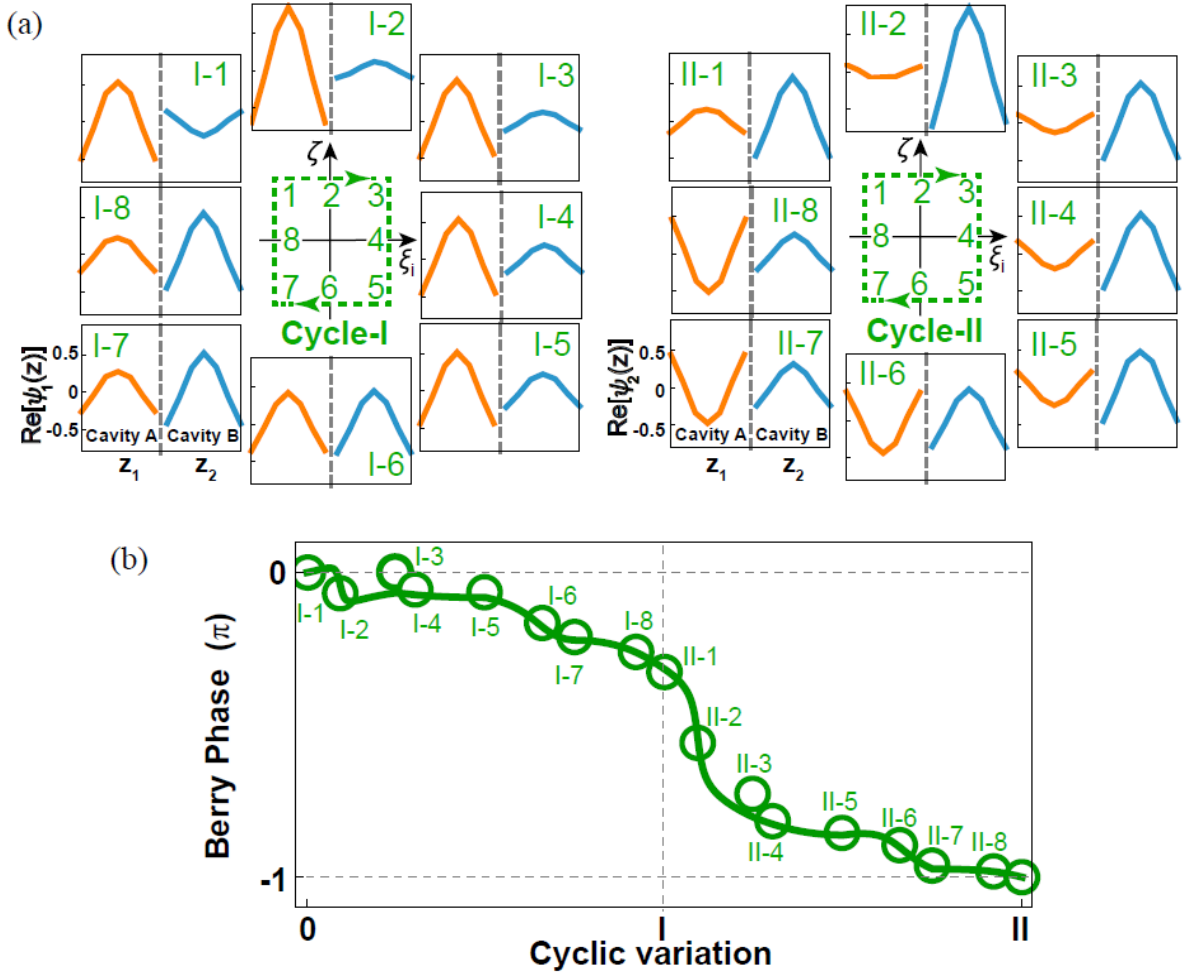

FIG. S6. (a) The measured eigenfunctions (real parts) along the Loop-A' encircling a single EP (Fig. 1(a), main text), which lies on $\xi_i\zeta$-plane with $\zeta = -0.24$. The orange and cyan lines respectively denote cavities A and B. The two eigenfunctions are out-of-phase at their starting point (Point I-1). State exchange occurs near Point I-2 and starting from Point I-3, the two cavities become in phase until the end of the first cycle. The second cycle begins with a state (II-1) that is the same as the state at I-8. States exchange again near II-2 and the two cavities are out-of-phase again and are restored upon two complete cycles, with a phase difference of $\pi$. (b) The corresponding Berry phase is $-\pi$. The markers are experimental results and the curves are from theory.

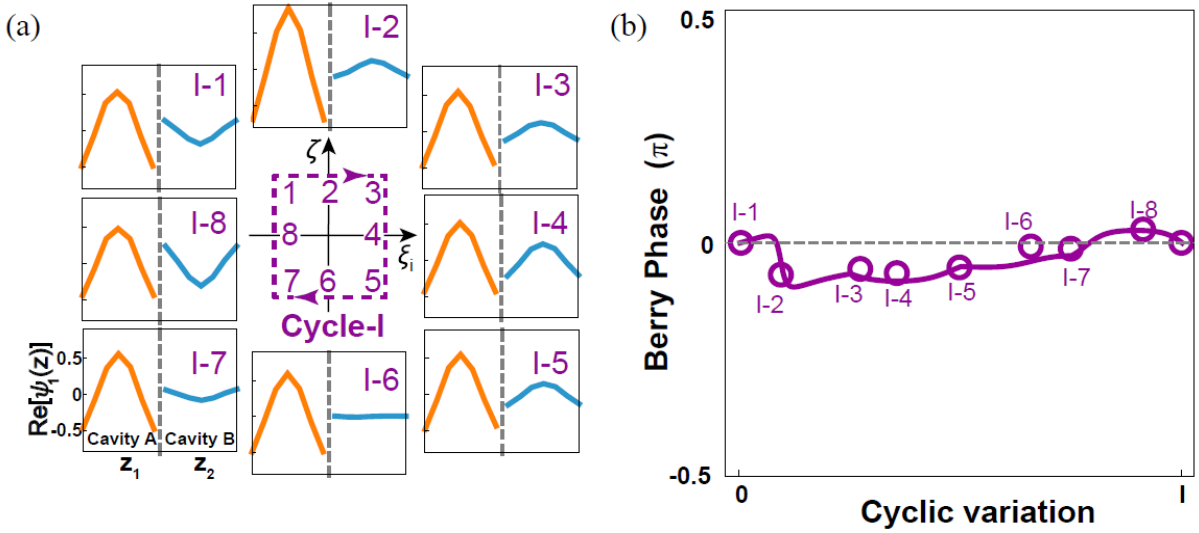

FIG. S7. (a) The measured eigenfunctions (real parts) along Loop-B' encircling the two EPs, which lies on $\xi_i\zeta$-plane with $\zeta = -0.24$ (Fig. 1(a), main text). The orange and cyan lines respectively denote cavities A and B. The two wavefunctions are out-of-phase at the starting point (Point I-1). The two states exchange twice near Points I-2 and I-6. The eigenstates are restored at the end of one cycle. (b) The corresponding Berry phase is 0. Markers are experimental results and curves are from theory.

## IX. Vertex EP as a chain point

As a topological invariant defined for complex eigenvalues, the DN offers a straightforward route to analyze the topology related to multiple exceptional points. To illustrate, we briefly show an example here. Consider a Hamiltonian slightly different from the Eq. (2) in the main text,

$$H_{EP} = \kappa_0 \begin{bmatrix} 0 & -1 + (\zeta_r + i\zeta_i)^2 \\ -1 + (\zeta_r + i\zeta_i)^2 & -2i(1 + \xi) \end{bmatrix}, \quad (21)$$

wherein $\kappa_0, \xi, \zeta_r, \zeta_i \in \mathbb{R}$. Compared with Eq. (1) in the main text, here the onsite parameter $i\xi$ is purely imaginary, but the hopping parameter is complex $\zeta_r + i\zeta_i$. Equation (21) generates two EPBs that kiss at their vertices, forming a chain point at $(\xi, \zeta_r, \zeta_i) = (0,0,0)$ (Fig. S8(a)). Using the DNs, it is very straightforward to analyze how this case is topologically different from the one in the main text. By encircling the EP chain point (red star) with the dashed green loop, we obtain a DN of $\mathcal{V}_{\zeta_r\zeta_i} = 2$, as shown in Fig. S8(d, e). In accordance to this non-trivial DN, Fig. S8(b) shows a point gap near the chain point instead of a line gap. Next, by noting that the dashed blue loop is homotopic to the green one (Fig. S8(a)), we can easily identify that the two EPs on the two arms of the EPB (red squares) must each have a DN of 1, which is confirmed in Fig. S8(c, d, f). The DNs clearly indicate that the EPB produced by Eq. (21) is topologically distinct from the one reported in our paper. In fact, the non-trivial DNs lead us to conclude that the loops in Fig. S8(a) are not homotopic to a constant loop, which already implies the existence of the second EPB and together with the first EPB produce the EP chain point. Without using DNs as a criterion, it will be much more difficult to make such a prediction and analysis.

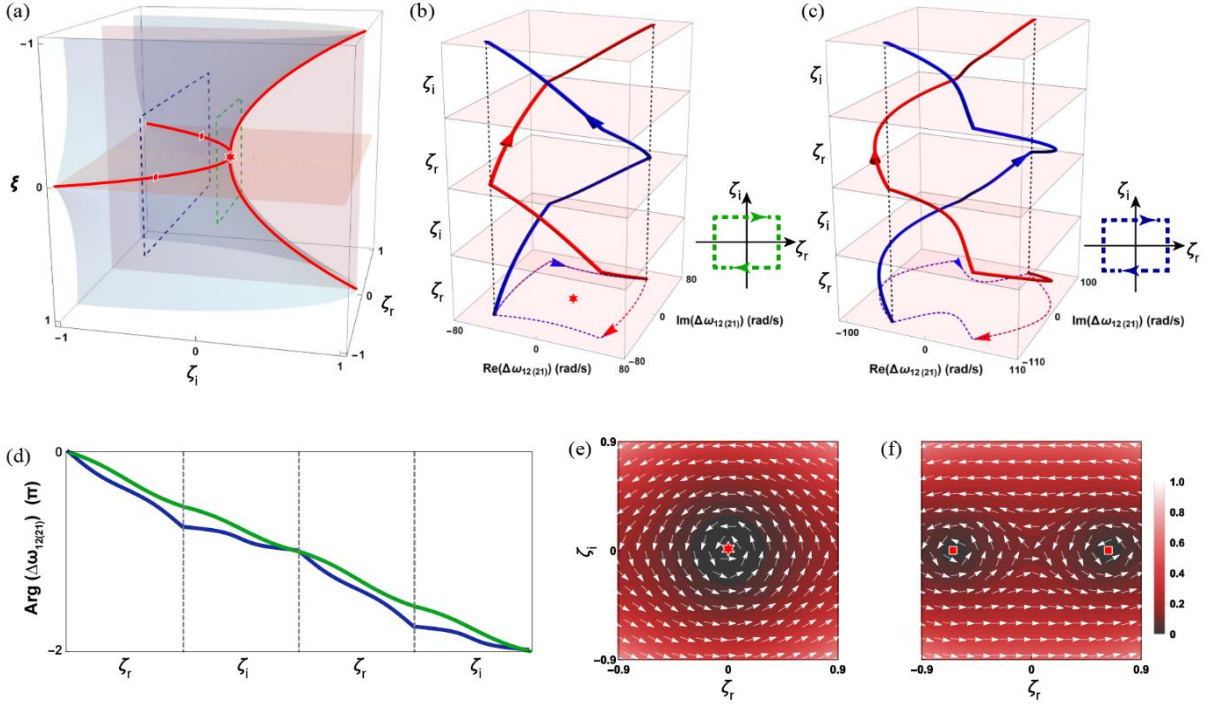

FIG. S8. (a) Eigenvalues of Eq. (21) give two exceptional parabolas forming a chain point (red star). The evolution of eigenvalue differences is plotted in (b, c) for the two loops shown in (a). Both results indicate that a point gap, which is different from the case discussed in the main text. (d) The evolution of eigenvalue differences $\Delta\omega_{12}$ along the dashed green loop encircling the chain point and the dashed blue loop encircling both arms of an exceptional parabola. Both yield the same DN $\mathcal{V}_{\zeta_r \zeta_i} = 2$. (e) and (f) show the discriminant fields on the two encircling planes in (a), respectively.

## X. Lists of parameters

Here, we present the parameters in our experiments. The second longitudinal mode resonates at $\omega_0 = 19,613 \text{ rad/s}$. The intrinsic loss of each cavity is $\gamma_0 = 83.5 \text{ rad/s}$. When $\zeta = 0$ (corresponding to opening six holes (two rows) in the middle ($z = 0$)), the hopping strength is $\kappa_0 = 48.5 \text{ rad/s}$. In Tables S1 – S4, we list the values of $\xi_r, \xi_i, \zeta$ along the four encircling paths that we have investigated (Fig. 1(a), main text). As mentioned, these parameters are obtained by performing a least-square fitting to the experimentally measured pressure response spectra using the Green's function. The fitted pressure responses have relative errors in the range of 0.024 – 0.086. Some examples of the fitted results are shown in Fig. S9, where good agreement with measured results is seen.

TABLE S1. The parameters on the solid green loop with $\zeta = 0$.

| Point # | $\xi_r$ | $\xi_i$ |
|---|---|---|
| 1 | 0.14 | -0.28 |
| 2 | 0.10 | 0.00 |
| 3 | 0.13 | 0.36 |
| 4 | -0.44 | 0.37 |
| 5 | -1.00 | 0.37 |
| 6 | -1.00 | 0.01 |
| 7 | -1.00 | -0.26 |
| 8 | -0.45 | -0.28 |

TABLE S2. The parameters on the solid purple loop with $\xi_r = 0$.

| Point # | $\zeta$ | $\xi_i$ |
|---|---|---|
| 1 | 0.65 | -0.50 |
| 2 | 0.66 | 0.00 |
| 3 | 0.66 | 0.36 |
| 4 | 0.12 | 0.35 |
| 5 | -0.68 | 0.36 |
| 6 | -0.65 | 0.00 |
| 7 | -0.66 | -0.48 |
| 8 | 0.51 | -0.48 |

TABLE S3. The parameters on the dashed green loop with $\xi_r = -0.24$.

| Point # | $\zeta$ | $\xi_i$ |
|---|---|---|
| 1 | 0.53 | -0.29 |
| 2 | 0.51 | 0.00 |
| 3 | 0.54 | 0.49 |
| 4 | 0.41 | 0.51 |
| 5 | 0.02 | 0.51 |
| 6 | 0.02 | 0.01 |
| 7 | 0.03 | -0.29 |
| 8 | 0.36 | -0.29 |

TABLE S4. The parameters on the dashed purple loop with $\xi_r = -0.24$.

| Point # | $\zeta$ | $\xi_i$ |
|---------|---------|---------|
| 1 | 0.53 | -0.29 |
| 2 | 0.52 | 0.02 |
| 3 | 0.54 | 0.49 |
| 4 | 0.00 | 0.51 |
| 5 | -0.72 | 0.51 |
| 6 | -0.72 | 0.00 |
| 7 | -0.72 | -0.29 |
| 8 | 0.05 | -0.29 |

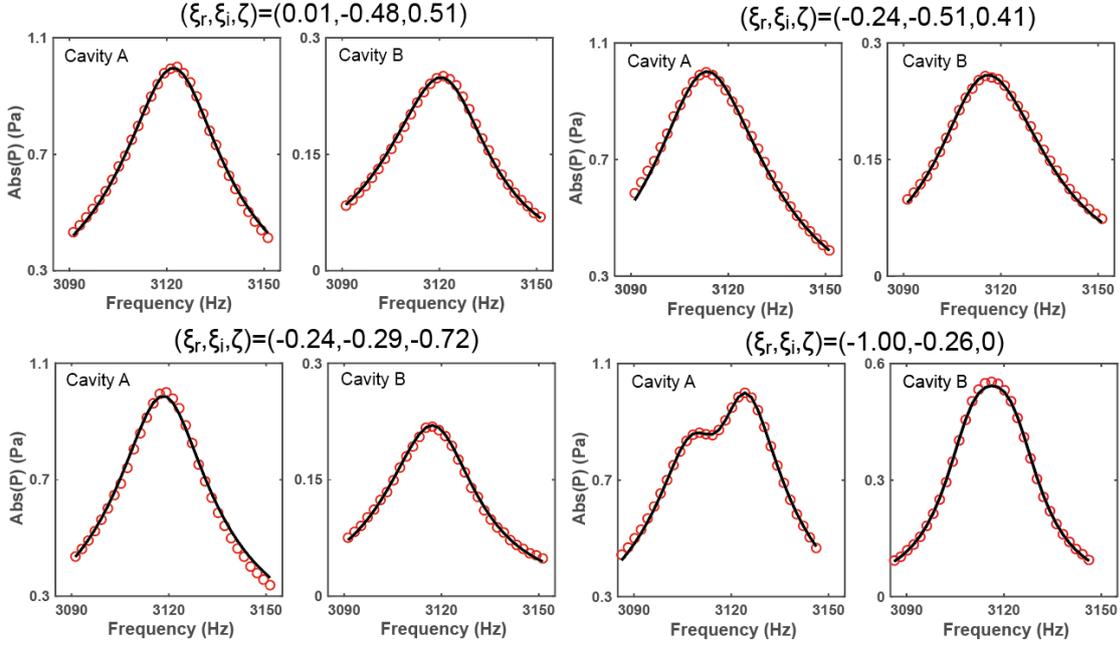

FIG. S9. A few examples of the measured (red open circles) and fitted (black curves) pressure response spectra. Good agreement between the two shows the validity of our fitting scheme.